\documentclass[12pt, draftclsnofoot, onecolumn]{IEEEtran}

\usepackage{cite}
\usepackage{amsmath,amssymb,amsfonts}
\usepackage{algorithmic}
\usepackage{graphicx}
\usepackage{epstopdf}
\usepackage{textcomp}
\usepackage{xcolor}
\usepackage{subfigure}
\usepackage{amsthm}
\usepackage{mathrsfs}

\usepackage{pdfpages}
\usepackage{colortbl}
\definecolor{mygray}{gray}{0.85}
\usepackage{array}
\usepackage{caption}
\usepackage{setspace}
\usepackage{diagbox}
\definecolor{mygray}{rgb}{0.96,0.99,0.96}
\definecolor{mypink}{rgb}{.99,.93,.85}
\definecolor{mycyan}{cmyk}{.2,0.04,0,0}
\usepackage[linesnumbered,boxed,ruled,commentsnumbered]{algorithm2e}
\usepackage{xcolor}  
\usepackage{colortbl,booktabs}
\usepackage{multirow}
\usepackage[ruled,linesnumbered]{algorithm2e}

{}
{}
{}

\newcolumntype{I}{!{\vrule width 1.25pt}}
\newlength\savedwidth

\newlength\savewidth
\newcommand\shline{\noalign{\global\savewidth\arrayrulewidth
		\global\arrayrulewidth 1.25pt}%
	\hline
	\noalign{\global\arrayrulewidth\savewidth}}

\usepackage{amsmath,bm}
\usepackage{cite}
\usepackage{amsmath,amssymb,amsfonts}
\usepackage{algorithmic}
\usepackage{amsthm}
\usepackage{graphicx}
\usepackage{textcomp}
\usepackage{xcolor}

\begin{document}

\title{\Huge Converged Reconfigurable Intelligent Surface \\and Mobile Edge Computing for\\ Space Information Networks
}

 \author{
\IEEEauthorblockN{Xuelin Cao, Bo~Yang,~\IEEEmembership{Member,~IEEE},
 Chongwen Huang,~\IEEEmembership{Member,~IEEE},\\
 Chau Yuen,~\IEEEmembership{Fellow,~IEEE},
 Yan Zhang,~\IEEEmembership{Fellow,~IEEE}, 
 Dusit Niyato,~\IEEEmembership{Fellow,~IEEE},
 and Zhu Han,~\IEEEmembership{Fellow,~IEEE}
 }
\thanks{X. Cao, B. Yang, and C. Yuen are with the Engineering Product Development Pillar, Singapore University of Technology and Design, Singapore. (e-mail: xuelin$\_$cao, bo$\_$yang, yuenchau@sutd.edu.sg).}%
\thanks{C. Huang is with the College of Information Science and Electronic Engineering, Zhejiang Provincial Key Lab of information processing, communication and networking, and International Joint Innovation Center, Zhejiang University, China. (e-mail: chongwenhuang@zju.edu.cn).}
\thanks{Y. Zhang is with the Department of Informatics, University of Oslo, Norway. (e-mail: yanzhang@ieee.org).}
\thanks{D. Niyato is with the School of Computer Science and Engineering, Nanyang Technological University, Singapore. (e-mail: dniyato@ntu.edu.sg).}
\thanks{Z. Han is with the Department of Electrical and Computer Engineering, University of Houston, USA. (e-mail: zhan2@uh.edu).}
}
\maketitle

\begin{abstract} Space information networks (SIN) are facing an ever-increasing thirst for high-speed and high-capacity seamless data transmission due to the integration of ground, air, and space communications. However, this imposes a new paradigm on the architecture design of the integrated SIN. Recently, reconfigurable intelligent surfaces (RISs) and mobile edge computing (MEC) are the most promising techniques, conceived to improve communication and computation capability by reconfiguring the wireless propagation environment and offloading. Hence, converging RISs and MEC in SIN is becoming an effort to reap the double benefits of computation and communication. In this article, we propose an RIS-assisted collaborative MEC architecture for SIN and discuss its implementation. Then we present its potential benefits, major challenges, and feasible applications. Subsequently, we study different cases to evaluate the system data rate and latency. Finally, we conclude with a list of open issues in this research area.
\end{abstract}


\IEEEpeerreviewmaketitle

\section{Introduction}
Space information networks (SIN) make it possible to provide seamless connectivity services to any object anywhere, such as urban areas and rural areas, even in isolated areas (e.g., desert, ocean, and mountain areas) not be easily reached by traditional networks. In recent years, SIN has become a promising network architecture that significantly extends the wireless coverage. It has also opened many new frontiers for network operators and service providers to deploy versatile and uninterrupted services on various application scenarios \cite{KXue, TDe, CZhang}. However, providing high-speed and high-capacity services in SIN poses severe challenges when integrating the ground, air, and space infrastructures. It becomes important for SIN to integrate with the emerging network technologies to adapt to this complicated communication environment \cite{NU, NKato, XZhu}.


Promisingly, advances in metamaterials have fuelled research in reconfigurable intelligent surfaces (RISs) for beneficially reconfiguring the wireless communication environment with the aid of a large array of inexpensive antennas. This new paradigm has been verified to bring several potential benefits for future wireless communications, such as coverage enhancement, data rate increase, and spectrum/energy efficiency improvement. Because of these potential benefits, RISs are eminently suitable for addressing various challenges of SIN by initiating the RIS-assisted ground and air communications, e.g., RIS-assisted multi-user systems for energy efficiency and latency \cite{CHuang, TBai, XCao1}, and RIS-assisted UAV communications \cite{XCao}. On another note, mobile edge computing (MEC) technology, thanks to its dense geographical distribution and comprehensive mobility support, improves the computational capability not only in densely populated areas but also in sparsely populated areas. Thus, MEC is conceived in a bid to fill the gap between the centralized cloud and mobile users, which becomes a promising paradigm for SIN to inspire the development of myriads of applications. Clearly, converging RISs and MEC in SIN has to be adopted fully exploiting the potential of communications and computations.

Several ongoing research activities show that SIN can benefit from the satellite MEC implementation, e.g., wireless converge enhancement, quality of service (QoS) improvement \cite{RXie, ZZhang}. In particular, the satellite MEC technologies are proposed to improve the QoS of SIN-based communication systems by using a dynamic network functions virtualization (NFV), and a cooperative computation offloading (CCO) model \cite{ZZhang}. Furthermore, by extending edge computing capabilities to the satellite, the multi-layer edge computing architecture and heterogeneous edge computing resource co-scheduling issues are further explored \cite{RXie}. Additionally, artificial intelligence (AI) techniques are investigated to address the optimization problems related to computation offloading \cite{NCheng1,BYang}. To further reduce the energy and delay cost of edge AI applications, a multiple-algorithm service model is proposed for energy-delay optimization \cite{WZhang}.

Utilizing RISs in different MEC platforms of SIN needs to deal with the following technical challenges: 1) how to perform offloading according to the capability of MEC platforms; 2) how to jointly schedule MEC platforms and RISs to benefit from each other; 3) how to improve the performance of MEC platforms by reconfiguring RISs. As a consequence, it is important to explore an effective SIN architecture to increase the system data rate and reduce the latency by converging RISs and MEC. The main contributions of this article are listed as follows.
\begin{itemize}
\item[-]  We propose an RIS-assisted collaborative MEC architecture for SIN. In this architecture, RISs and MEC platforms are integrated in SIN to improve the capability of communications and computations.
\item[-]  We design different offloading schemes for the different MEC platforms. Accordingly, we further present an implementation strategy for the proposed RIS-assisted collaborative MEC.
\item[-]  We discuss the benefits, challenges, applications, and services of RIS-assisted collaborative MEC. Then, we investigate three cases to evaluate the performance of the proposed architecture.
\end{itemize}

The rest of this article is organized as follows. Section II proposes an RIS-assisted collaborative MEC architecture with emphasizing its offloading schemes and implementation strategy. Section III discuss the benefits, challenges, and applications of the proposed RIS-assisted collaborative MEC. The case studies are presented in Section IV. Finally, Section V concludes the paper and poses several open issues.


\begin{figure*}
\small
	\centering{\includegraphics[width=6.3in, height=3in]{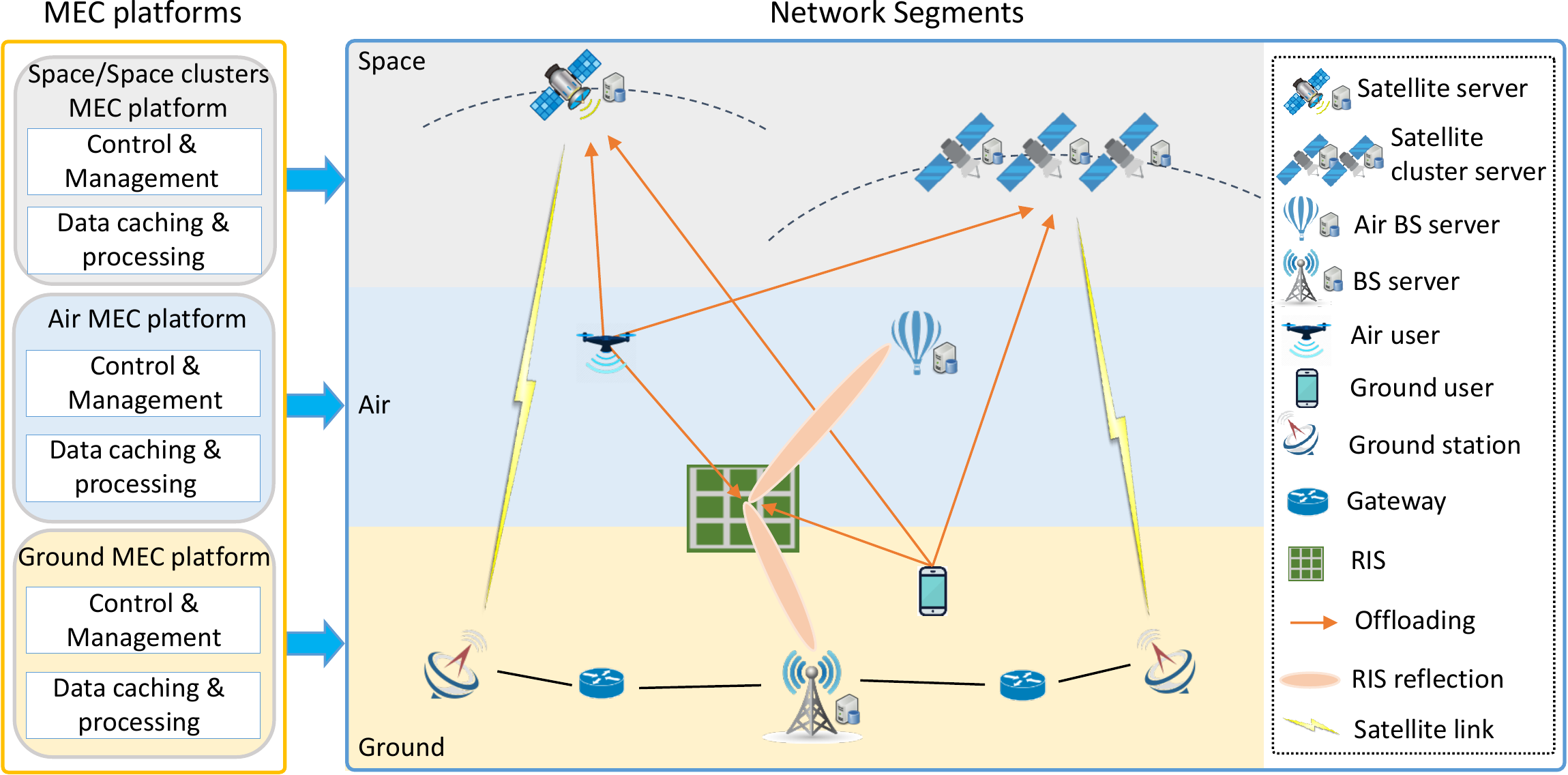}}
	\caption{RIS-assisted collaborative MEC architecture in space information networks.}
	\label{A1}
\end{figure*}

\section{RIS-Assisted Collaborative MEC Architecture in Space Information Networks}
In this section, we present an RIS-assisted collaborative MEC architecture for SIN, as shown in Fig. \ref{A1}. We then illustrate its network segments, MEC platforms, and implementation, respectively.

\subsection{Network Segments}\label{subsec1}
Different from the existing SIN system architectures, four segments are illustrated when MEC platforms and RISs are integrated into SIN.  
\par\textbf{RIS-Assisted Ground Segment}. The RIS-assisted ground segment equipped with MEC serves various ground users by leveraging cellular networks, wireless local area networks (WLAN), mobile ad hoc networks (MANET), and low-power wide-area networks (LPWAN).  The RIS-assisted ground segment can provide high data rate and low latency services via RISs, although its wireless coverage is limited.
\par\textbf{RIS-Assisted Air Segment}. The RIS-assisted air segment consists of low and high-altitude platforms (LAP/HAP) equipped with MEC, which employs UAVs/aircrafts as carriers to implement information acquisition/processing and provide broadband wireless communications via RISs. The air segment and ground segment mutually complementing and reinforcing each other. Compared with the RIS-assisted ground segment, the RIS-assisted air segment is easy to deploy, while its wireless coverage can be significantly enhanced at low cost, while sometimes its link not stable enough.
\par\textbf{Space Segment}. The space segment is composed of the geosynchronous equatorial orbit (GEO), medium earth orbit (MEO), and dense low earth orbit (LEO) satellites equipped with MEC, constellations, and the corresponding ground segment infrastructures like the ground station and core network. Satellite communications should satisfy the unique demands of large area coverage and high-speed propagation.
\par\textbf{RIS-Assisted Cross-Segment}. Cross-segment concerns the integration of ground, air, and space. It involves Ground-to-Air (G2A) segments, Ground-to-Space (G2S) segments, Air-to-Space (A2S) segments, Ground-Air-Space (GAS) segments, where GAS combined G2A with A2S is extending of G2S. Since the soaring development of the RIS-assisted ground and air segments, G2A and GAS can support flexible RIS-assisted communications and provide various QoS requirements. With the assistance of MEC and RISs, these four types can complement each other and immensely enrich communication resources.

\begin{figure*}[t]
\small
	\centering{\includegraphics[width=4.2in, height=2.9in]{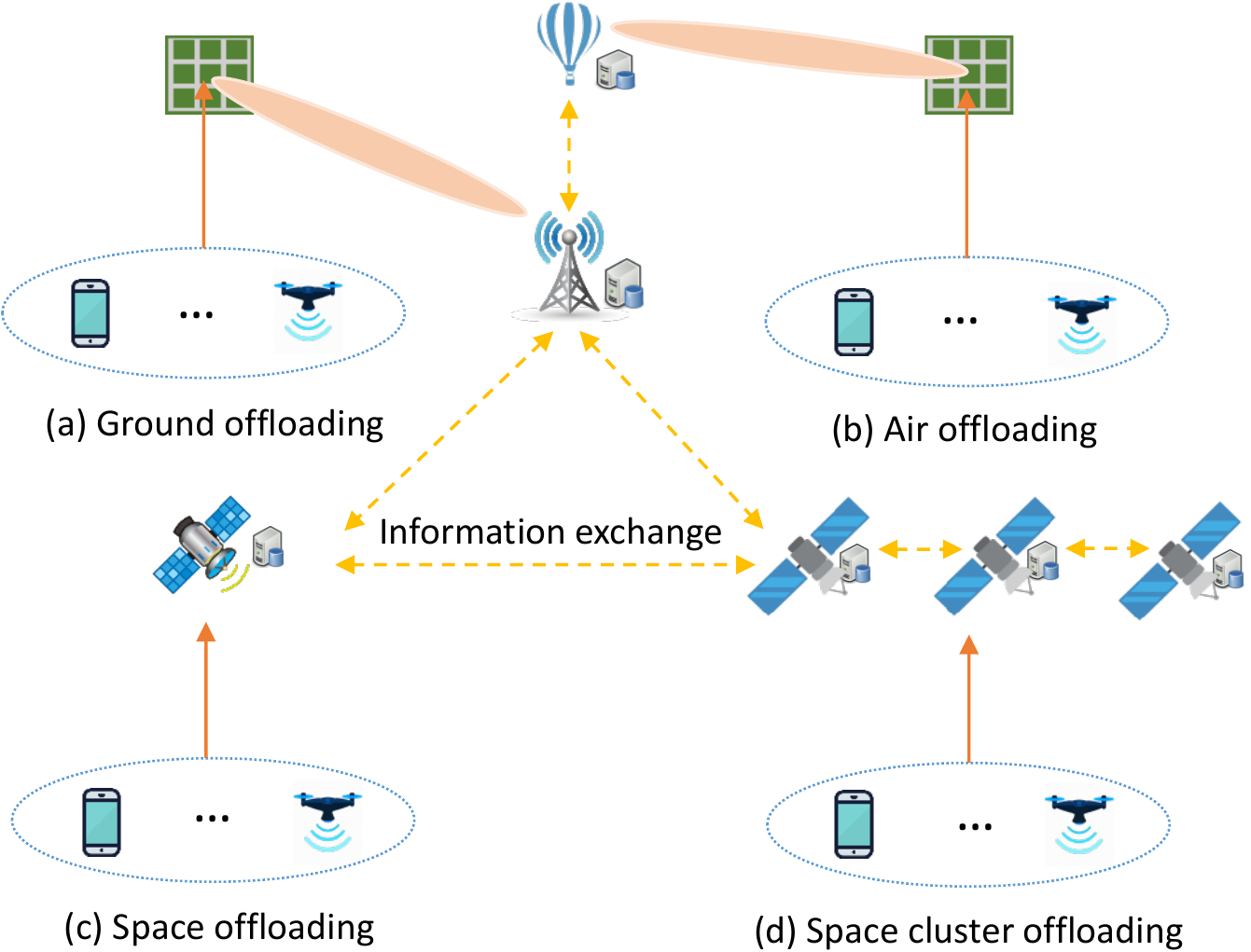}}
	\caption{Offloading in RIS-assisted collaborative MEC.}
	\label{A2}
\end{figure*}

\subsection{MEC Platforms and Offloading}\label{subsec2}

As shown in Fig. \ref{A1}, each segment of SIN maintains an MEC platform for the control, management, data caching, and task processing. Hence, the offloading scheme of each segment must be carefully designed considering its unique features.

\par\textbf{Ground MEC Platform}. MEC deployed in the ground base station, named the ground MEC platform, has strong computing capabilities. Therefore, the ground MEC platform can be employed to implement the most abundant RIS-assisted ground offloading, as shown in Fig. \ref{A2} (a). RIS-assisted ground offloading can support high data rates and high throughput services for individual users. When ground or air users offload their computation task to the ground MEC platform via RIS, the low latency can be achieved since the benefits of RISs can be fully exploited in short-distance Non-Line-of-Sight (NLOS) transmission.

\par\textbf{Air MEC Platform}. Air MEC platform is defined as a MEC platform enabled on the air base station, and it has fewer computation resources due to the power constraint. Thus it is deemed an effective complement of the ground MEC platform. As shown in Fig. \ref{A2} (b), it allows the ground or air users to implement the RIS-assisted air offloading when the RIS-assisted ground offloading cannot meet their QoS demands or it is unavailable. The RIS-assisted air offloading is characterized by wide coverage and flexible deployment. It can process the ground or air users' offloading at low-cost by using RISs, thereby accelerating their task processing through collaboration with the RIS-assisted ground offloading.

 \begin{figure*}
	\centering{\includegraphics[width=6.5in, height=3in]{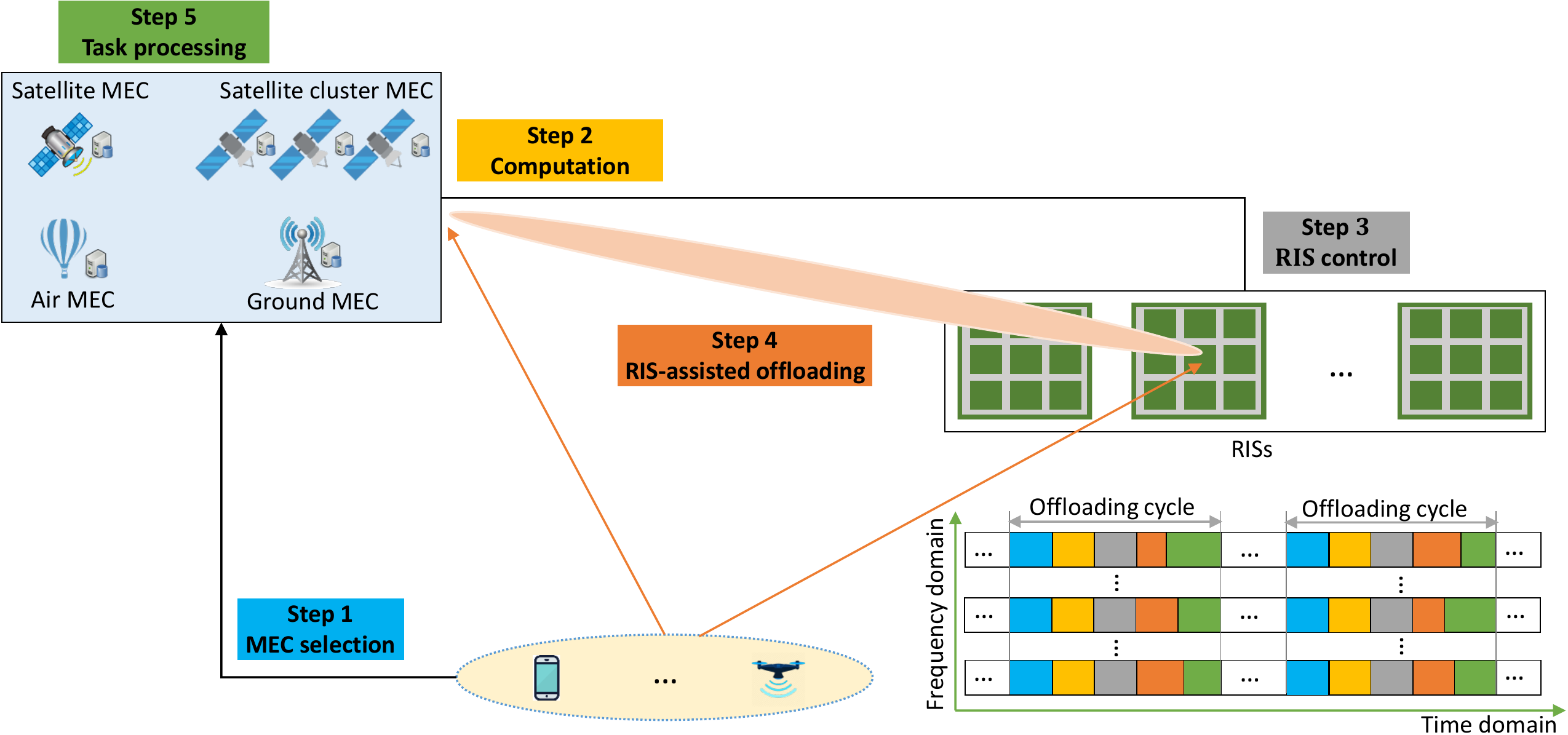}}
	\caption{Implementation of RIS-assisted collaborative MEC.}
	\label{A3}
\end{figure*}

\par\textbf{Space MEC Platform}. The MEC platform extended to the LEO satellite is deemed to a space MEC platform. It tackles the computation task via the satellite networks, as shown in Fig. \ref{A2} (c). It provides the space offloading when the RIS-assisted ground and air offloading cannot meet the demands of task computation and processing. In the space offloading \cite{RXie}, the space MEC platform has the most extensive communication coverage and broadcasting capability to serve numerous users anywhere. It also allows users to offload their computation task to the satellite, thus speeding up the task processing of users without the aid of ground and air segments. Additionally, by flexible collaboration with the RIS-assisted ground and air offloading, the space offloading may be used to alleviate intensive computation of the ground or air MEC platform. Note that RISs are unavailable in the space due to the long distance.

\par\textbf{Space Cluster MEC Platform}. Space cluster MEC platform is an extent of the space MEC platform, i.e., a group of non-identical LEO satellites that can cooperate with each other constitutes a space cluster. It provides more flexible offloading services and more computation resources. In one space cluster, an LEO satellite equipped with an MEC platform has its specific role, and it can be dynamically combined to provide space cluster offloading, as shown in Fig. \ref{A2} (d). In this offloading, the ground or air users offload their computation task to the air space cluster platform to process when the space MEC platform's computation resource is unavailable or insufficient. By intra-cluster collaboration, the space cluster MEC platform can provide sufficient computing resources and better services.

In the proposed offloading architecture, the ground or air users are service consumers, and the ground, air, space, space cluster MEC platforms are the leading service providers that can exchange information with each other. Since the limitation of energy and mobility, the air, space, and space cluster MEC platforms can be regarded as lightweight platforms. Therefore, once the task processing of one MEC platform is heavy, the computation task can be sent to the data center via the backbone network and Internet.

\subsection{Implementation} \label{subsec3}

In order to explain the proposed architecture, we present the implementation of the RIS-assisted collaborative MEC, as shown in Fig. \ref{A3}. In this architecture, the air and space MEC platforms collaborate with the ground MEC platform to support the RIS-assisted task offloading. Explicitly, the offloading request and MEC selection are triggered by users, while the resource allocation and the RIS configuration are completed collaboratively at MEC platforms. In particular, as a user generates tasks, the user first handles its tasks leveraging the local computation resource. Once the local computation resource is insufficient or even no, the user needs to offload its tasks to an MEC platform in an offloading cycle. The implementation of the RIS-assisted collaborative MEC is presented as follows.


\par \textbf{Step 1: MEC Selection}. Each user selects one MEC platform to send its offloading request according to the status of MEC platforms and its tolerable delay. Fig. \ref{A3} shows that the user sends its offloading request to the ground MEC platform if $flag_g==1$ and $delay_g\leq\delta$, otherwise it judges $\{flag_a, delay_a\}$, $\{flag_s, delay_s\}$, and $\{flag_{sc}, delay_{sc}\}$ sequentially to select the corresponding MEC platform. Here, $flag_g==1$ and $delay_g\leq\delta$ denote that the ground MEC platform is available and the delay to the ground MEC platform is lower than or equal to the latency threshold ($\delta$), respectively.

\par \textbf{Step 2: Computation}. After receiving offloading requests from users, MEC platforms share the request information, optimize the resources, and feed back the optimization results to users, where computations include the following three aspects.
\begin{itemize}
\item[-] \textit{Offloading Scheduling}. According to its affordable computation capability, the selected MEC platform schedules the task offloading for users.

\item[-] \textit{Resource Allocation}. To guarantee QoS requirements, the selected MEC platform allocates the spectrum, power, and RIS resources for users.

\item[-] \textit{RIS Reconfiguration}. The RIS reconfiguration parameters are optimized at the selected MEC platform to support the task offloading. Note that the RIS reconfiguration can be affected by the number of RISs, the size of each RIS, the number of users, and the propagation links between users and MEC platforms.

\end{itemize}

\par \textbf{Step 3: RIS Control}. According to the calculated RIS resource allocation and the RIS reconfiguration, the selected MEC platform controls the RIS to assist the user's offloading. Note that RISs have marginal effects when offloading to the space and space cluster MEC platforms due to the long distance. 

\par \textbf{Step 4: RIS-Assisted Offloading}. Once the user receives the feedback of the selected MEC platform, it offloads its tasks to the selected MEC platform with the aid of RISs. 

\par \textbf{Step 5: Task Processing}. The selected MEC platform processes the offloaded tasks from users. After finishing task processing, the MEC platform feeds back the results to users via RISs.

With this implementation, the computation load for users is small since the user only needs to calculate its task processing delay that is required to offload its tasks to the different MEC platforms. If users' computation tasks are related to the network environment or the state of servers, the task offloading and MEC selection are intertwined in an SIN setup. In this case, MEC selection should be addressed by considering both the user and MEC platform sides. Additionally, some preliminary works on MEC technologies, including MEC integrated with the ground, air, space, and cross-segment, are summarized in Table \ref{TB} to show the differences to this article.

\begin{table*}[t] 
\newcommand{\tabincell}[2]{\begin{tabular}{@{}#1@{}}#2\end{tabular}}
		\small
		\centering
			\renewcommand{\arraystretch}{1.2}
			\captionsetup{font={small}} 
			\caption{\scshape Comparison of Different Segments in SIN} 
			\label{TB}
			\footnotesize
			\centering  
			\begin{tabular}{|m{0.08\textwidth}<{\centering}|m{0.07\textwidth}<{\centering}|m{0.08\textwidth}<{\centering}| m{0.08\textwidth}<{\centering}|m{0.04\textwidth}<{\centering}|m{0.04\textwidth}<{\centering}|m{0.16\textwidth}<{\centering}|m{0.16\textwidth}<{\centering}|m{0.06\textwidth}<{\centering}| }  
				\shline
			    \rowcolor{mycyan} \textbf{Segments} & \textbf{Entities} & \textbf{Altitude} &\textbf{Round-trip} & \textbf{MEC} & \textbf{RIS} & \textbf{Advantage} & \textbf{Disadvantage} & \textbf{Ref.}\\
				\shline  
				Ground & Cellular / WLAN / MANET/ LPWAN \ \  & N.A. & Lowest & $\mathbf{\surd}$ & $\mathbf{\surd}$ & Rich resources, \ \ \ \ \ \ \ \ \ \ \ \ \ \ high data rate,\ \ \ \ \ \ \ \ \ \ \ \ \ \ \ \ \ high throughput,\ \ \ \ \ \ \ \ \ and low latency. & Limited coverage, vulnerable to disaster, \ \ \ \ \ and high mobility. & \cite{TBai, XCao1, CHuang, BYang, WZhang}\\ 
			     \hline
			    Air & HAP / LAP \ \ \  & up to 30\ \ km & Medium & $\mathbf{\surd}$ & $\mathbf{\surd}$ & Wide coverage,\ \ \ \ \ \ \ \ \ \
low cost, \ \ \ \ \ \ \ \ \ \ \ \ \ \ \ \ low latency,\ \ \ \ \ \ \ \ \ \ \ \ \ \ \ \  and flexible deployment. & Less capacity, \ \ \ \ \ \ \ \ \ \ high mobility, \ \ \ \ \ \ \ \ \ \ \ and link instability. & \cite{XCao, NCheng1}\\ 
			    \hline
			    Space & GEO / MEO / LEO - satellite  & 160-35786 km & Highest & $\mathbf{\surd}$ & $\mathbf{\times}$ & Broadcast/multicast, large coverage, \ \ \ \ \ \ \ \ \ \ \ and rapid commercialization. & Long propagation delay,\ \
limited capacity,\ \ \ \ \ \ least flexibility,\ \ \ \ \ \ \ \ \ \ and costly. & \cite{KXue, TDe, CZhang, ZZhang} \\ 
			    \hline
			    Cross-Segment & G2A / G2S  / A2S / GAS  \ \  & up to 35786 km & Flexible & $\mathbf{\surd}$ & $\mathbf{\surd}$ & Rich resources, \ \ \ \ \ \ \ \ \ \ \ \ collaborative,\ \ \ \ \ \ \ \ \ \ \ \ \ \ \ flexible services,\ \ \ \ \ \ \ \ \ \ \ \ \ \ \ and high data rate. & Complexity,\ \ \ \ \ \ \ \ \ \ \ \ \ \ \ \ \ \ \ \ \ \ \ \ link instability,\ \ \ \ \ \ \ \ \ \ \ \ \ \ extra overhead, \ \ \ \ \ \ \ \ \ \ \ \ \ \ \ \ \ \ \ and high mobility. & \cite{NU, NKato, XZhu, RXie}\\ 
			    \shline
			\end{tabular}  
	\end{table*}

\section{Benefits, Challenges, and Applications} 

With the proposed RIS-assisted collaborative MEC architecture for SIN, the following potential benefits, major challenges, promising applications and services are investigated.

\subsection{Potential Benefits}
The communication, computation, and RIS benefits in SIN brought by the proposed RIS-assisted collaborative MEC architecture are discussed.    

\textbf{Communication Benefits}. Such benefits are mainly manifested in communication coverage, data rate/delay, and security.

\begin{itemize}
\item[-] \textit{Coverage Enhancement}. The communication coverage may be enhanced due to the double action of dense deployment of collaborative MEC platforms and RISs. On the one hand, by collaboration among the ground, air, and space MEC platforms, an offloading area of over 1 million km$^2$ can be approximately covered. On the other hand, by leveraging software-controlled RISs in the ground and air, the coverage of MEC can be furthered enhanced in a small area.
\item[-] \textit{Data Rate and Latency Improvement}. Deploying different MEC platforms in different segments and extending their services to the programmed wireless environments can improve the system data rate and reduce the latency. Moreover, the performance of MEC platforms can be further improved by scheduling and reconfiguring RISs.
\item[-]\textit{Security and Privacy}. Deploying the extensive MEC platforms and RISs in SIN requires a large number of services, which will result in complicated security issues. Enabling the different transmissions via RISs on different MEC platforms can provide a high security and privacy. 
\end{itemize}

\textbf{Computation Benefits}. Such benefits to users are mainly manifested in computing delay, energy, and service capacity.

\begin{itemize}
\item[-] \textit{Computing Delay Reduction}. Collaboration among different MEC platforms can expand computing capabilities, thus assisting users to speed up their resource allocation and task processing via RISs when the local computation resource is scarce or cannot meet the QoS requirements. For this reason, the MEC platform, with the aid of RISs and other MEC platforms, can provide various services for delay-sensitive and compute-intensive applications.
\item[-]\textit{Energy Saving}. The energy-saving of MEC platforms mainly benefits from the following two aspects. On the one hand, task computing can be offloaded to different MEC platforms by leveraging RISs; on the other hand, different MEC platforms can be provided differentiated computing models for various services to reduce energy consumption. Such energy-saving is extremely important for the space segment that is powered by solar energy.  
\item[-]\textit{Service Capacity Enlargement}. With the aids of RISs and MEC platforms, more extensive users can be served by different MEC platforms to process their task offloading, thus improving the computation capability of MEC platforms.   
\end{itemize}

\textbf{RIS Benefits}. In the conventional single-layer MEC system, e.g., the ground MEC system, although RISs can improve the quality of wireless links. However, the communication gain benefited from RISs may be decreased or even offset when the MEC server is overloaded. At this time, the offloading latency is mainly restricted by the computational time at the MEC platform. We call this the `straggler effect'. To avoid the straggler effect, converging RISs and cross-segment MEC platforms in SIN is fully exploited to strengthen the offloading performance. In return, by appropriately allocating RISs to support different MEC platforms, the superiority of the RISs can be fully explored.

\begin{figure*}[t]
\small
 \centering
          	\subfigure[Total delay vs. the number of ground users.]{
    		\begin{minipage}[b]{0.465\textwidth}
    		\centering{\includegraphics[width=1\textwidth]{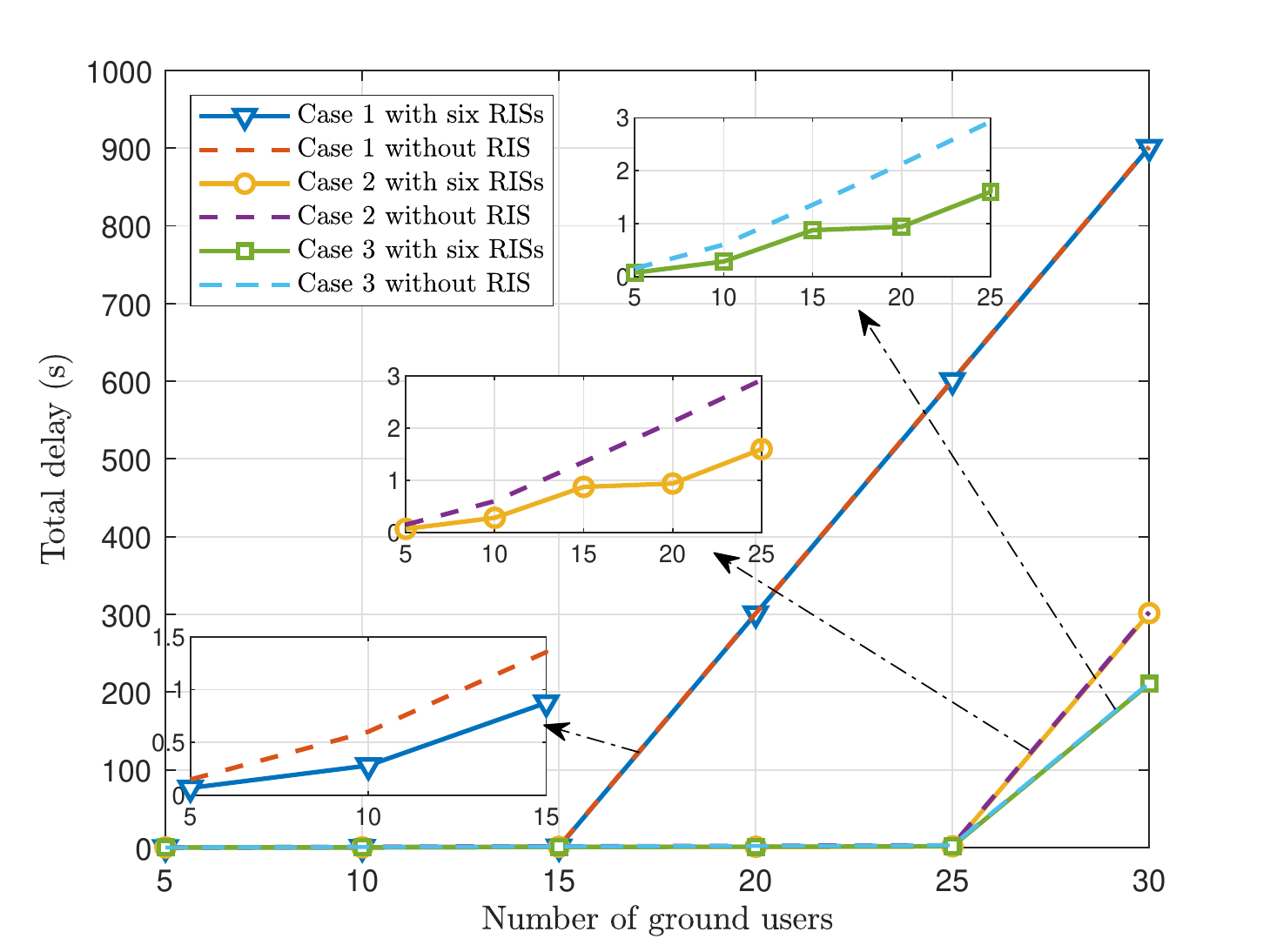}}
    		\end{minipage}
    		\label{S1a}
    	}
    	\hfill
       	    \subfigure[Total computing delay and data rate vs. the number of ground BS servers.]{
    		\begin{minipage}[b]{0.465\textwidth}
    		\centering{\includegraphics[width=1\textwidth]{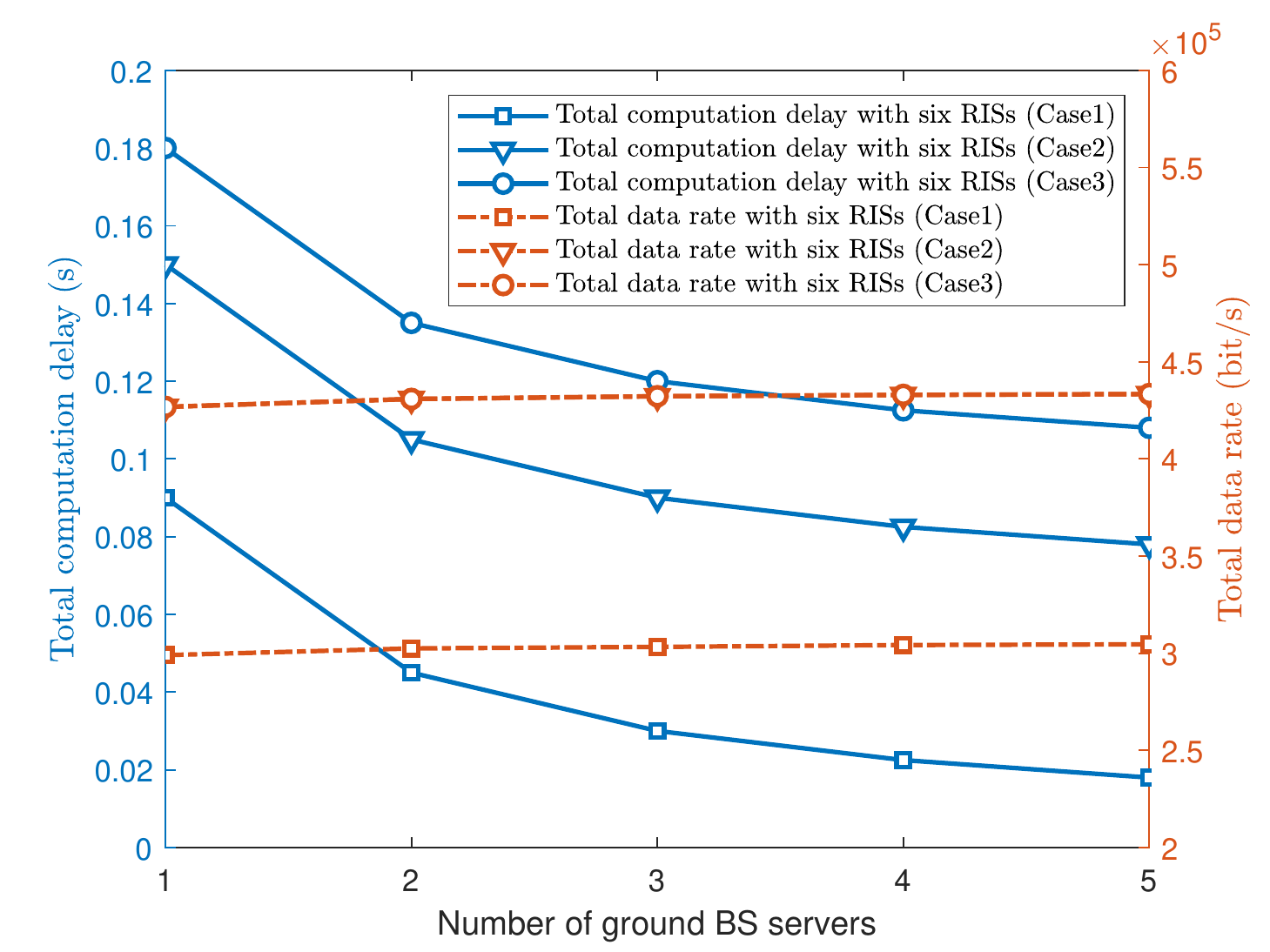}}
    		\end{minipage}
    		\label{S1b}
    	}   
    	\caption{Total delay, total computing delay, or total data rate vs. the number of ground users or BS servers, respectively.} 
    	\label{S1}
    \end{figure*} 

\subsection{Major Challenges}
\par Converging RISs and MEC platforms in SIN introduces several new challenges.
\par\textbf{Collaboration}. The collaboration of different SIN segments and different MEC platforms may affect the implementation of RIS-assisted collaborative MEC architecture. Thus, achieving collaboration with each other to improve the efficiency of communications and computations is still a vital challenge.
\par\textbf{Computation}. The complicated resource allocation combined with extensive task processing and RISs reconfiguration may bring an enormous of computation at the different MEC platforms. How to improve the computation capability of MEC platforms is critical for ensuring efficient resource utilization, low latency, and reliability.
\par\textbf{Autonomous Management}. The dynamic SIN imposes the new challenges on autonomous management of RIS-assisted collaborative MEC architecture, such as self-sustaining networks, adaptive MEC service migration, and flexible implementation as RISs are leveraged to assist offloading.   
\par\textbf{Access Control}. Supporting massive users offload their tasks to the different MEC platforms via RISs is based on efficient access control of all involved entities, such as the radio access of users, MEC platforms, and RISs. Thus, medium access control (MAC) design for RIS-assisted collaborative MEC architecture is critical.
\par\textbf{Signal Processing}. To effectively improve the RIS benefits on cross-segment MEC platforms, the signal processing involving RIS channel estimation, RIS interference suppression, and RIS modulation is becoming more challenging. 
\par\textbf{Standardization}. Since the ground networks, air networks, and satellite networks may belong to different Internet service providers, converging RISs and MEC in SIN needs to consider the standardized operation, maintenance, and management of different MEC platforms.

\subsection{Applications and Services}
In view of the above-mentioned potential benefits and challenges, RIS-assisted collaborative MEC architecture for SIN will be more applicable in the following application scenarios.    
\par\textbf{Sixth-Generation (6G)}. It can be applied in 6G to provide ubiquitous computing among local devices and MEC platforms, thus achieving accurate sensing, monitoring, and control.  
\par\textbf{mmWave/TeraHertz}. It may be considered in mmWave/ TeraHertz communications due to their high propagation attenuations and molecular absorptions. The RIS-assisted collaborative MEC makes it feasible for enlarging transmission distance and coverage range in mmWave/TeraHertz frequencies.  
\par\textbf{Transportation Service}. It can be widely applied in the intelligent transportation system (e.g., terrestrial, aerial, and maritime) to provide safe driving services for high mobility vehicles, drones, and vessels, even providing these services in the isolated areas. 
\par\textbf{Emergency Service}. It can be applied in the military field to provide emergency services in war zones, where cellular communications may be destroyed or terminated, such as remote monitoring and strike. It also provides public healthcare services in disaster areas, such as remote diagnosis and tracking of Covid-19, remote surgery, and so on. 
\par\textbf{Hologram Telepresence}. It can be used in hologram telepresence for image processing and transmitting that involves people and/or objects at a remote location, thus providing genuinely immersive virtual reality/augmented reality (VR/AR) services.

\section{Case Studies}
In this section, we first give scenario setting and then investigate three cases to evaluate the proposed RIS-assisted collaborative MEC architecture.
\subsection{Scenario Setting}

We consider a scenario that consists of a ground BS server, an air BS server, an LEO satellite server, $6$ RISs with $256$ RIS-elements each, and $30$ ground users, where the location of the ground BS server, the air BS server, and the LEO satellite server are set to be [0.2 0.2 0] km, [0.1 0.1 0.3] km, and [0.1 0.1 160] km, respectively. Explicitly, 3 RISs for the ground segment and 3 RISs for the air segment. The computation capacity of the ground BS server, the air BS server, and the satellite server are 1.5 G, 1 G, and 0.5 G, respectively. The delay threshold of each ground user, $\delta$ is set to be 60 s. It is assumed that the ground and air channels are NLoS links, and the space channels are line-of-sight (LoS) links. The power dissipated at each user is 30 dBm, the bandwidth is 10 MHz, and the noise power is -94dBm.  

When the ground segment cannot meet QoS requirements effectively, collaboration with the air and space segments is required. For instance, as a user cannot be served by the ground MEC platform in a congested area or isolated area, it possibly leverages the air MEC platform or space MEC platform to continue its task offloading. In general, the following three cases can be considered:

%
%

\begin{itemize}
\item[-] \textbf{Case 1}. Ground users only offload their tasks to the ground BS server.   
\item[-] \textbf{Case 2}. Ground users can offload their tasks to the ground BS server or the air BS server.
\item[-] \textbf{Case 3}. Ground users can offload their tasks to the ground BS server, the air BS server, or the satellite server.
\end{itemize}

\subsection{Results Evaluation}

\textbf{Total Delay Reduction}. The total delay is defined as the summation of computing delay and communication delay of all ground users. Fig. \ref{S1a} shows that the total delay in three cases, where Case 3 is the lowest, and Case 1 is the highest as the number of ground users increases. This is because that the proposed RIS-assisted collaborative MEC architecture can process the ground users' tasks by providing air and space computation resources. Also, the total delay can be reduced as RISs are introduced into the three cases. Note that the total delay in each case increases quickly as the number of ground users increases due to the communication delay of the increased unserved ground users. Additionally, Fig. \ref{S1b} shows that the total computation delay in each case decreases as the number of ground BS servers increases.

\textbf{Total Data Rate Improvement}. The total data rate is defined as the summation of all ground users' data rates. In Fig. \ref{S1b}, the total data rate in each case remains almost unchanged as the number of ground BS servers increases, and the reason is that the communication capability of ground networks remains unchanged. Moreover, it can be seen from Fig. 5 that the total data rates of Case 2 and Case 3 are higher than that of Case 1 due to the following reasons. The ground users can offload their tasks to the air BS server or the satellite server, thereby improving the total data rate. Compared to Case 2, the total data rate improvement in Case 3 is not obvious due to the long propagation delay. Additionally, a higher total data rate can be achieved in each case when more RISs are deployed, and then a noticeable improvement benefited from RISs can be observed in all three cases.

\begin{figure}
\small
	\centering{\includegraphics[width=0.5\textwidth]{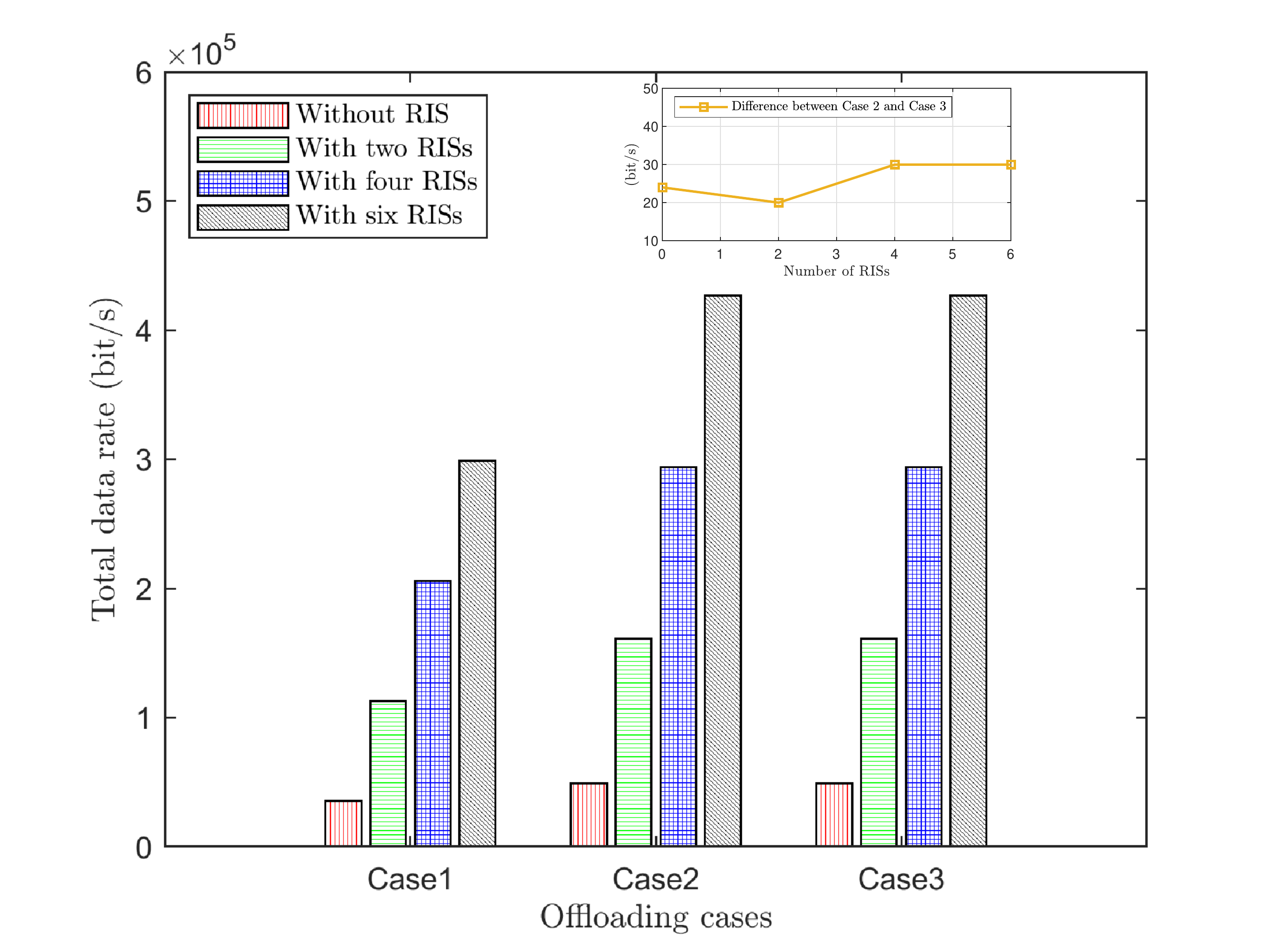}}
	\caption{Total data rate vs. the number of RISs.}
	\label{S2}
\end{figure}

\section{Conclusions}\label{sec7}
In conclusion, existing RISs and MEC technologies have been introduced into SIN to improve the data rate and latency. We first proposed an RIS-assisted collaborative MEC architecture for SIN, where the ground, air, space segments were integrated into one system. In particular, different MEC platforms can provide different offloading for users by cooperation. Then we presented an implementation of the proposed RIS-assisted collaborative MEC with highlighting its five steps. On this basis, we discussed their potential benefits, major challenges, promising applications and services. By studying three cases, it was shown that the proposed RIS-assisted collaborative MEC architecture for SIN can benefit from the total data rate and delay.

We foresee some open issues in RIS-assisted collaborative MEC for SIN. For example, a community effort is required for adopting advanced AI, mmWave/Thz, and software-defined networking techniques for SIN. Additionally, security and privacy-preserving should be taken into consideration as a large number of network entities are involved. Promisingly, SIN architecture designs integrated with next-generation technologies are becoming an exciting area for new research.



\vspace{-10 mm}
\end{document}